\documentclass[11pt,twoside]{article}
\usepackage{asp2004}
\usepackage{psfig}
\usepackage{epsf}
\usepackage{graphics}
\usepackage{lscape}
\usepackage[dvips]{graphicx}

\def\edcomment#1{\iffalse\marginpar{\raggedright\sl#1\/}\else\relax\fi}
\reversemarginpar

\begin{document}
\title{Modelling the progenitors of Core-Collapse Supernovae}
\author{J. J. Eldridge and C. A. Tout}
\affil{University of Cambridge, Institute of Astronomy, The Observatories, Madingley Road, Cambridge, CB3 0HA, United Kingdom.}

\begin{abstract}
We present details of our investigation of the progenitors to
core-collapse supernovae. We discuss observations and the theory of
the lowest-mass stars to explode as supernovae.
\end{abstract}
\thispagestyle{plain}

\section{Introduction}
A star's evolution to SN depends on four factors, initial mass,
initial metallicity, mass loss and the inclusion of extra mixing above
that from convection described by mixing-length theory.  Mass loss is
the most uncertain with many groups using their own preferred
rates. We have experimented with a number of prescriptions to compare
them and see which fit observations such as Wolf-Rayet (WR) star type
ratios and SNe observations. Detailed results can be found in
\citet{et1} and \citet{eldridge1}. Here we present an example of
our results and discuss in particular the lowest mass stars that
explode as SNe.

\section{Modelling SN Progenitors}
We use the Cambridge stellar evolution code written by
\citet{E71} and most recently updated by
\citet{etopac}.  We include convective overshooting
as described by \citet{SPE97}.  Our
mass-loss prescription is similar to that described by
\citet{et1}. However we make two alterations.  First we scale all mass
loss, including WR mass loss, with initial metallicity such that
$\dot{M}(Z)=\dot{M}(Z_{\odot})(Z/Z_{\odot})^{0.5}$. Second we
supplement the pre-WR evolution with the rates of \citet{VKL2001}.

We have created various grids of models to study the features of SN
progenitors.  These are useful to
observational searches for progenitors now underway
\citep{S03,VD03}.  This is similar to the work of
\citet{H03}, \citet{vanb03} and \citet{izzy} but we have studied
many different mass-loss schemes and examined details of the
progenitors such as luminosity, temperature, radius and final
mass.  Our range of initial masses used is $M_{\rm initial}=5$ to $200
\, M_{\odot}$.  Our metallicity range is $Z=0.001$ to~$0.05$,
with $Z_\odot = 0.02$.

\begin{figure}[!t]
\begin{center}
\includegraphics[height=90mm,angle=0]{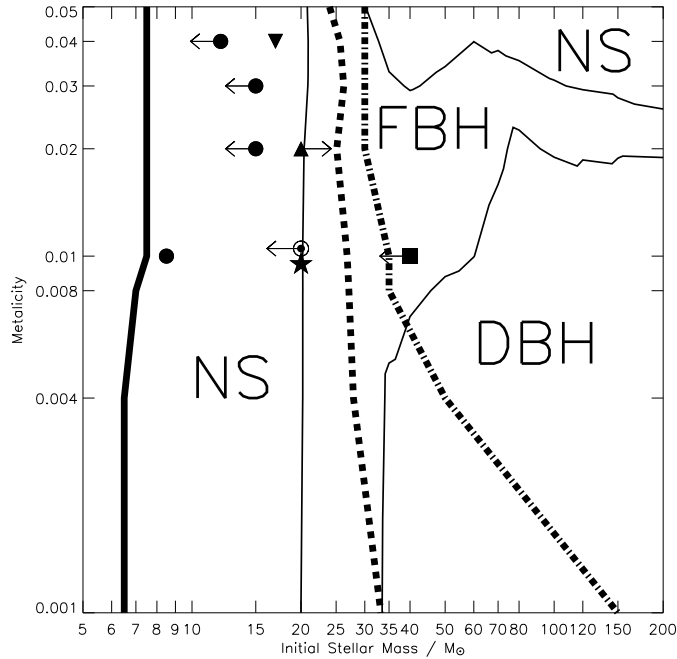}
\end{center}
\caption{Supernova properties -- thin contours separate regions where
different remnants form in the SNe. NSs are neutron stars,
FBHs are black holes formed by fallback on to a black hole and DBHs are
black holes that form directly. The thick solid line is
the minimum mass for SNe. The dashed line separates IIP SNe
to its left from IIL SNe.  The dash-dotted line is
the minimum mass for type Ibc SNe for which all hydrogen has been
removed from the progenitors which are Wolf-Rayet stars. The points
are observed SNe \citep{S03}: the filled circles
are IIP SNe, the square a Ic SN, the star a IIpec SN, the
triangle a IIn SN and the inverted triangle a IIb SN.}
\label{figA}
\end{figure}

Figure~\ref{figA} is an example of our results that match observations
such as observed WR type and SN type ratios reasonably well.  Most
importantly we agree with the observations of the SN progenitors
described by Smartt et al. (2003; 2004).  The IIP progenitors are in
the correct region, as is the sole Ic.  The IIpec and IIb SNe are
SN1987A and SN1993J both of which have probably been affected by
binary star evolution \citep{podsi92,Maund04}. The IIn SN, 1997bs, is
thought not to be a SN at all \citep{VD00}.  The remaining one, 1980K,
has an observed mass too low to agree with our models.  It is also
likely to be the result of binary evolution but we discuss another
possibility below.

\section{Low-Mass SN Progenitors}

\begin{figure}[!t]
\begin{center}
\includegraphics[height=105mm,angle=270]{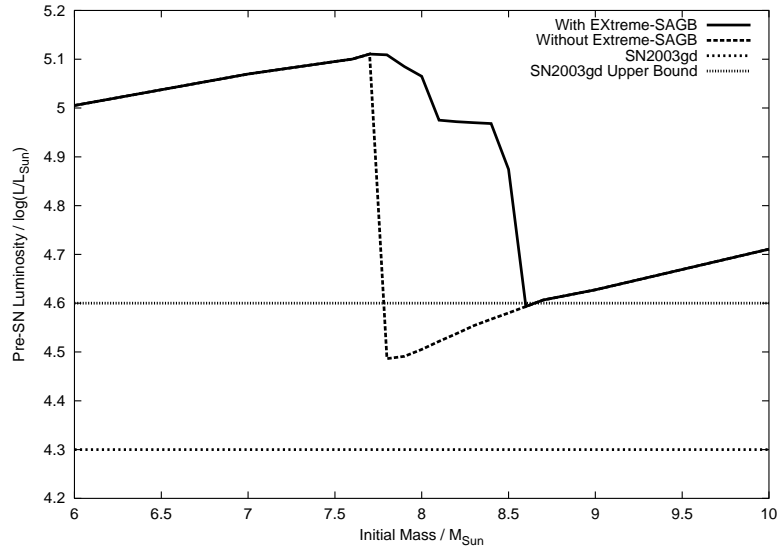}
\end{center}
\caption{The luminosity of SN progenitors.
The solid line includes the extreme SAGB stars while the
dashed line does not.
The horizontal lines are the observed luminosity of the progenitor of
2003gd and its upper error bar.}
\label{figB}
\end{figure}

SN2003gd was the first red giant progenitor observed and is the lowest
mass progenitor found.  In figure \ref{figB} we plot the
luminosity of our models against initial mass together with the
limits of the luminosity for 2003gd, $\log(L/L_{\odot})=4.3 \pm 0.3$
\citep{SJM03}.

In the figure we see two populations of progenitors, those of
high and low luminosities.  The high-luminosity stars are AGB stars
in which second dredge-up has occurred.  This is the
deepening of the convective envelope after shell helium is established on the AGB and the H-burning shell is extinguished.
This carries hydrogen to a
hotter environment near the helium burning shell and the luminosity of
the star increases.  With the hydrogen and helium burning shells in
close proximity thermal pulses occur. Low-mass AGB
stars have carbon oxygen (CO) cores and lose their envelopes during
thermal pulses so they do not undergo SNe.  More massive stars ($M_{\rm
initial} > 7M_{\odot}$) ignite carbon before dredge-up and therefore
have oxygen neon (ONe) cores during the thermal pulses, these are
Super-AGB stars. Again many of these lose their envelopes and
become white dwarfs.  However the most massive Super-AGB stars have core
close to, or above, the Chandrasekhar limit ($M_{\rm Ch}$) at second
dredge-up. This means that there is a little time to lose the
envelope before core collapse. In an ONe core collapse is
triggered when the central density is high enough for electron capture
on to $^{24}$Mg.  This removes the electron degeneracy pressure that
supports the core and it collapses to a neutron star. Such stars have
been modelled by \citet{IBEN1} and \citet{IBEN2}.  We find similar
behaviour but at different initial masses owing to our new opacity tables
and convective overshooting.

At an initial mass of $7.5 \, M_{\odot}$ we find that the core mass
after second dredge-up is about $1.38 \, M_{\odot}$.  The core does
not grow but achieves the conditions for core-collapse about 10,000 years
later.  Any luminosity increase at second
dredge-up would increase the mass-loss rate. The low initial mass of
the envelope before second dredge-up means little
would remain at the time of the SN.  This would exclude a plateau phase in the light
curve and so these would be IIL~SNe.  Such
stars therefore fit the mass limits of 1980K, its change in mass-loss
history and its SN type.  However this type of progenitor to IIL~SNe
would be rare and we can expect them to be dominated by binary
systems.

When the initial mass is greater than about $7.8 \, M_{\odot}$ we find
the CO core exceeds $M_{\rm Ch}$ before second dredge-up.  An
intershell convection zone forms and reduces the mass of the core.  In
these stars the time from second dredge-up to core collapse is between
10 and~100$\,$yr.  We find that our models only just fit within the
error bars of the observation of 2003gd. However if we take the
luminosity before second dredge-up in the most extreme Super-AGB stars
we find that the region which overlaps with observations is larger.
This could mean that these extreme Super-AGB stars do not occur.  To
prevent second dredge-up either the helium or the CO core must
grow or we must inhibit convection, because inhibiting convection
delays second dredge-up so it does not occur before
core-collapse.  Convection itself is very uncertain and models
are simple. However in future we can refine our theory and use
observations of SN progenitors similar to that of SN2003gd to provide
a limit on the nature of convection in stellar interiors.

\section{Future Work}
Our future plans include exploring the effect of convection on
lowest-mass SN progenitors and looking for low-luminosity progenitors
in models of binary stars.

\begin{acknowledgements}
JJE would like to thank PPARC for his studentship and Fitzwilliam College for his scholarship. CAT would like to thank Churchill College for his fellowship.
\end{acknowledgements}

\end{document}